\documentclass[conference]{IEEEtran}

\usepackage{epsfig,cite}
\usepackage[tbtags]{amsmath}
\DeclareGraphicsExtensions{.eps}
\usepackage[font=footnotesize]{subfig}
\usepackage{dblfloatfix}
\usepackage[para]{threeparttable}
\usepackage{soul}
\usepackage{graphicx}
\usepackage{caption}
\usepackage{cite}
\usepackage{multirow}
\usepackage{algorithm}
\usepackage{algorithmic}

\usepackage{fancyhdr}

\begin{document}

\title{Minimax Design and Order Estimation of FIR Filters for Extending the Bandwidth of ADCs}

\author{
\IEEEauthorblockN{Yinan Wang\IEEEauthorrefmark{1}\IEEEauthorrefmark{2}, 
H\aa kan Johansson\IEEEauthorrefmark{2}, Hui Xu\IEEEauthorrefmark{1}, and Jietao Diao\IEEEauthorrefmark{1}}
\IEEEauthorblockA{\IEEEauthorrefmark{1}College of Electronic Science and Engineering, National University of Defense Technology, Changsha 410073, China}
\IEEEauthorblockA{\IEEEauthorrefmark{2}Division of Communication Systems, Department of Electrical Engineering, Link\"{o}ping University, SE-581 83, Sweden\\ Email: wangyinan@nudt.edu.cn, hakanj@isy.liu.se, xuhui@nudt.edu.cn}
}

\maketitle

\thispagestyle{fancy}
\cfoot{\scriptsize{Copyright 2016 IEEE. Published in the 2016 IEEE International Symposium on Circuits and Systems (ISCAS 2016), scheduled for 22-25 May 2016 in Montreal, Canada. Personal use of this material is permitted. However, permission to reprint/republish this material for advertising or promotional purposes or for creating new collective works for resale or redistribution to servers or lists, or to reuse any copyrighted component of this work in other works, must be obtained from the IEEE. Contact: Manager, Copyrights and Permissions / IEEE Service Center / 445 Hoes Lane / P.O. Box 1331 /Piscataway, NJ 08855-1331, USA. Telephone: + Intl. 908-562-3966.}}
\renewcommand{\headrulewidth}{0pt}

\begin{abstract}
The bandwidth of the sampling systems, especially for time-interleaved analog-to-digital converters, needs to be extended along with the rapid increase of the sampling rate. A digitally assisted technique becomes a feasible approach to extend the analog bandwidth, as it is impractical to implement the extension in analog circuits. This paper derives accurate order estimation formulas for the bandwidth extension filter, which is designed in the minimax sense with the ripple constraints as the design criteria. The derived filter order estimation is significant in evaluating the computational complexity from the viewpoint of the top-level system design. Moreover, with the proposed order estimates, one can conveniently obtain the minimal order that satisfies the given ripple constraints, which contributes to reducing the design time. Both the performance of the extension filter and its order estimation are illustrated and demonstrated through simulation examples. 

\end{abstract}


\section{Introduction}
\label{sec:intro}
\IEEEPARstart{A}{nalog-to-digital} converters (ADCs) are generally required to perform with the flat frequency response up to the Nyquist band in many applications \cite{Dallet_1998}. It is however getting difficult for high-speed ADCs to satisfy this requirement due to either the front-end pre-amplifier or the sample-and-hold circuits in the converters. Especially for time-interleaved sampling systems, where the sampling rate is increased by a factor that equals the number of the interleaved channels, the effective bandwidth is however not proportionally increased as the sampling rate but restricted by the sub-ADCs \cite{Poulton_2003,SPDevices_2011,ApisSys_2012}. Increasing the bandwidth appropriately can improve the measurement performance\cite{Pickerd_2005,Agilent_2004,Lecroy_2010}. 
Some oscilloscope companies (e.g., Tektronix Inc, Agilent Inc, and Lecroy Inc) have developed this function into their instruments but gave no detailed design procedure\cite{Pickerd_2005,Agilent_2004,Lecroy_2010}. In \cite{Hars_2003}, the authors presented a gain compensation method to improve the magnitude response flatness based on several selected frequencies, however no approach for finding the minimal order was proposed. In summary, there is a lack of systematic design approaches with an accurate estimation of the computational complexity for the bandwidth extension of ADCs. 
  
This paper derives order estimation formulas of the minimax-designed finite-impulse-response (FIR) filter for extending the bandwidth of ADCs. The first advantage of the derived order estimation is that one can evaluate the computation complexity accurately, which is meaningful at the top-level design of the overall digital signal processing systems. Furthermore, one can substantially reduce the effort for searching the minimal order that satisfies the ripple requirements, since the estimates can offer a good initial value for the minimal order. Comprehensive simulations are presented to validate the performance of the derived order estimation formulas.

The rest of the paper is organized as follows. Section \ref{sec:prob_stat} presents the design problem for bandwidth extension and the minimax design. Section \ref{sec:ord_est} presents the filter order estimation by curve fitting. In Section \ref{sec:simexa}, the minimax-designed extension filter and its order estimation are verified through simulation examples. Section \ref{sec:conclusions} concludes this paper.

\section{Problem Statement and Minimax Design}
\label{sec:prob_stat}
\begin{figure}[t]
  \centerline{\includegraphics[width=6.8cm]{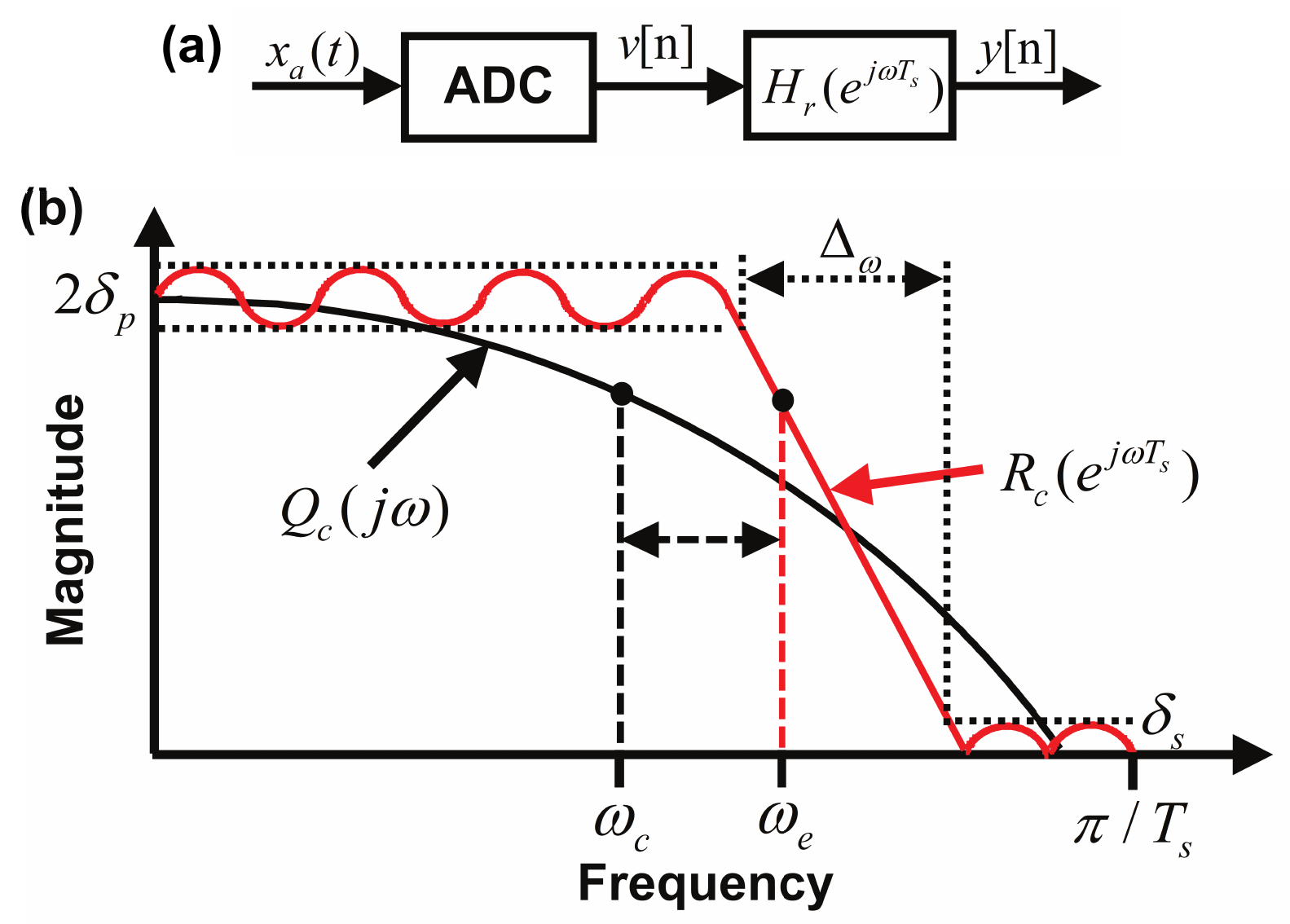}}  
  \caption{\footnotesize{(a) Principle of bandwidth extension using a digitally assisted technique. (b) Specification for the extension filter in the minimax sense.}}
  \label{fig:basandspe}
\end{figure}

Figure \ref{fig:basandspe}(a) illustrates the principle of bandwidth extension using a digitally assisted method. 
Assume that the analog input signal $x_a(t)$ is bandlimited to $\omega_e < \pi/T_s$, i.e., its Fourier transform satisfies $X_a(j\omega)=0,~|\omega| > \omega_e$, where the sampling period is $T_s$. The frequency response of the ADC is denoted as $Q_c(j \omega)$ with a $-3$~dB cutoff angular frequency of $\omega_c$, which satisfies $ \omega_c \leq \omega_e < \pi/T_s$ (see the model in \eqref{eq:1st_rc}). Sampling the continuous-time signal $x_a(t)$ at uniform time instances will generate the output sequence as $v[n]$, the spectrum of which is shaped by $Q_c(j \omega)$. Thus, in the frequency domain, we have
\begin{equation}
\label{perrec}
V\left(e^{j\omega T_s}\right) = \frac{1}{T_s} X_a\left(j\omega\right) Q_c\left(j\omega\right),\quad \omega T_s \in [-\pi,\pi],
\end{equation}
where $V\left(e^{j\omega T_s}\right)$ represents the discrete-time Fourier transform (DTFT) of $v[n]$. It is apparent that the original input signal is deteriorated by the non-ideal frequency response of the ADC. In order to enhance the flatness within the desired passband frequency region $[0,\omega_e T_s]$ as well as the capability for suppressing the distortion and noise outside the passband, a digital filter $h_r[n]$ is employed to equalize the ADC's frequency response. The frequency response of $h_r[n]$ with an order of $N$ is denoted as $H_r \left(e^{j \omega T_s}\right) = \sum^{N}_{n=0} h_r[n] e^{-j\omega T_s n}$. Therefore the equalized output response $Q_e(j \omega)$ through the expansion filter is ideally to have a unity gain in the extended passband and infinite attenuation in the remaining stopband, thus we have\footnote{Here, the desired passband response is $e^{-j \omega T_s \frac{N}{2}}$ instead of unity to take the delay of the extension filter into account.}
\begin{equation}
Q_e(j \omega)= 
\begin{cases}
e^{-j \omega T_s \frac{N}{2}} & \omega T_s \in [0,\omega_e T_s]\\
0   & \omega T_s \in (\omega_e T_s,\pi].
\end{cases}
\label{qe}
\end{equation} 
In practice, one can only approximate the ideal response, and in this paper we approximate it in the minimax sense.

\subsection{Minimax Design}
\label{subsec:mini}


Figure \ref{fig:basandspe}(b) shows the specification for the bandwidth extension filter designed in the minimax sense. Here, $\delta_p$ and $\delta_s$ represent the ripples of the passband and the stopband, respectively. To enable a practical minimax design, an additional transition band is introduced as $\Delta_\omega$. Thus we have the specification of the extension filter as
\begin{align}
\label{spec}
&\left| R_c\left(e^{j\omega T_s}\right) - Q_e(j \omega) \right| \leq \delta_p, \quad  \omega T_s \in [0,\omega_e T_s]& \notag \\
&\left| R_c\left(e^{j\omega T_s}\right) \right| \leq \delta_s, \quad  \omega T_s \in [\omega_e T_s + \Delta_\omega T_s,\pi],&
\end{align}
where $R_c\left(e^{j\omega T_s}\right)$ is given by
\begin{equation}
\label{rec_res}
R_c\left(e^{j\omega T_s}\right) = H_r \left(e^{j \omega T_s}\right) Q_c\left(j\omega\right).
\end{equation}

In order to meet the given specification in \eqref{spec}, the extension filter is here designed by solving the following approximation problem: For a given filter order $N$, find the filter coefficient $h_r[n]$ and $\delta$ so as to $\textbf{minimize} ~ \delta$
\begin{align}
\textbf{subject to} \quad \left| E\left( j \omega T_s\right) \right| < \delta,
\label{minimize}
\end{align}
on $\omega T_s \in [0,\omega_e T_s] \cup [\omega_e T_s + \Delta_\omega T_s,\pi]$, where $ E\left( j \omega T_s\right)$ is the weighted error function as given by
\begin{equation}
\label{error_fun}
E\left( j \omega T_s\right) = W(\omega T_s) \left[ R_c\left(e^{j\omega T_s}\right) - Q_e(j \omega) \right] 
\end{equation}
and
\begin{equation}
\label{wei_fun}
W(\omega T_s) = 
\begin{cases}
1 & \omega T_s \in [0,\omega_e T_s]\\
\delta_p/\delta_s   & \omega T_s \in [\omega_e T_s + \Delta_\omega T_s,\pi].
\end{cases}
\end{equation}

It is worth noting that this approximation problem is a convex problem, which means that one can obtain the globally optimal solution in the minimax sense. The corresponding finite-dimensional linear programming problem is solved by using standard optimization methods \cite{Nash_96}, and in this paper we utilize the general-purpose \textit{fminimax} in Matlab as well as the real-rotation theorem \cite{Parks_87}, which states that minimizing $|E(j \omega T_s)|$ is equivalent to minimizing $\Re \{E(j \omega T_s)e^{j\Theta}\}$, $\forall~\Theta \in [0,2\pi]$. The optimization problem is then solved with discretized $\omega T_s$ and $\Theta$. Typically, about 200-500 discrete grid points of $\omega T_s$ and 10-20 grid points of $\Theta$ are dense enough to satisfy the requirements. 

\section{Filter Order Estimation}
\label{sec:ord_est}
\begin{table*}[tb]
\renewcommand{\arraystretch}{1.3}
\caption{Estimated Parameters for $\Upsilon$ and $\Gamma$}
\label{tab:estpara}
\centering
\begin{threeparttable}[b]
\begin{tabular}{|c|c|c|c|c|c|c|c|c|c|}
    \hline
    \multirow{2}{*}{Region}   & \multicolumn{4}{c|}{$P$ values for $\Upsilon$} & \multicolumn{5}{c|}{$Q$ values for $\Gamma$} \\
    \cline{2-10}
       & $P_1$ &  $P_2$ & $P_3$ & $P_4$ & $Q_1$ & $Q_2$ & $Q_3$ & $Q_4$ & $Q_5$ \\
    \hline
    \hline
    Region 1 ($W_r \geq 1$)    & $0.9155$ &  $1.1199$  & $-0.0027$ & $0.0098$ & $-0.1682$ & $0.5913$ & $2.0607$ & $11.1035$ & $-6.115$ \\
    \hline
    Region 2 ($W_r < 1$)    & $1.2041$ &  $1.2962$  & $-0.0019$ & $0.0174$ & $-0.1023$ & $0.9368$ & $2.8292$ & $11.7762$ & $-8.725$ \\
    \hline
\end{tabular}
\end{threeparttable}
\end{table*}

From the practical perspective, it is desired to utilize the minimal filter order that can meet the given constraints on the ripples of the passband and the stopband. For this purpose, we derive the order estimates for the extension filter in this section.
Before giving the details of the filter order estimation, it is noted that the ADC's frequency response is modeled by a first-order RC circuit, which has been indicated as a reasonable assumption for many applications \cite{Van_2003,Tsai_06,Louwsma_2008}. Then we have
\begin{equation}
\label{eq:1st_rc}
Q_c(j \omega) = \frac{1}{1+j \omega \frac{1}{2\pi f_c}},
\end{equation}
where $f_c = \omega_c / 2\pi$ represents the $3$-dB cutoff frequency. Furthermore, we define the weighting ratio and the extension ratio respectively as 
\begin{equation}
\label{weiraoext}
W_r= \delta_p/\delta_s \quad \text{and} \quad \alpha = \omega_e/\omega_c.
\end{equation}

The filter order estimates are approached in two steps, which are presented as follows.
In the first step, we determine the basic form of the filter order estimation through numerous designs obtained according to the minimax sense in Section \ref{subsec:mini} with various transition bandwidths, weighting ratios, as well as expansion ratios. According to the numerous design examples, we have observed the basic form as
\begin{equation}
\label{ordest}
N_{est} = -\frac{\log_{10}(P_\delta)}{\Upsilon (\Delta_\omega,W_r)} + \Gamma (\Delta_\omega,W_r,\alpha),
\end{equation} 
where the order estimation is dependent on the product of the ripples $P_\delta = \delta_p \delta_s$ and the functions $\Upsilon$ as well as $\Gamma$. Further, it is observed that $\Upsilon$ is mainly dependent on $\Delta_\omega$ and $W_r$, and thus denoted as $\Upsilon(\Delta_\omega,W_r)$, whereas $\Gamma(\Delta_\omega,W_r,\alpha)$ represents a function of $\Delta_\omega$, $W_r$, and $\alpha$ as well.


Before giving the detailed procedure of the second step, it is necessary to discuss the reasonable range of values for $P_\delta$, $\Delta_\omega$, $W_r$, and $\alpha$. Considering a common acquisition system or ADC, the possible extended passband $\omega_e T_s$ is typically located within $[0.6\pi,0.9\pi]$, and $\omega_c T_s \leq \omega_e T_s$. Thus we have the rational expansion ratio as $\alpha \in [1,1.5]$. For the weighting ratio, it is typically satisfied with the ripple magnitude ranging from $-100$~dBc to $-20$~dBc (i.e., we have $\delta_p, \delta_s \in [0.00001,0.1]$) \cite{Rabiner_1975,Ichige_00}.\footnote{Although very small $\delta_p$ may be of less interest in actual sampling systems, it is included here for completeness.} Therefore the product and the weighting ratio of the ripples are within $[10^{-10},10^{-2}]$ and $[10^{-4},10^{4}]$, respectively. Furthermore, we focus on the transition bandwidth within $\Delta_\omega T_s \in [0.05\pi, 0.15\pi]$. In the future we will address more narrow transition band, which typically requires additional care \cite{Ichige_00}.\footnote{More narrow transition bandwidths (smaller than $0.05\pi$) are not generally utilized in practical applications, since it substantially increases the computational complexity. Furthermore, it is worth noting that one has to use much more complicated estimate functions for more narrow transition bands \cite{Ichige_00}.}

Under the above assumptions, we can get the specific estimation functions for $\Upsilon(\Delta_\omega,W_r)$ and $\Gamma(\Delta_\omega,W_r,\alpha)$ that can be determined through curve fitting by discretizing $P_\delta$, $\Delta_\omega$, $W_r$, and $\alpha$ in their respective range. Various criteria can be employed for such curve fitting. Here we solve the curve fitting problem in the minimax sense as well. Therefore, in the fitting procedure, the function form and its parameters are determined by minimizing the maximum deviation between the estimated filter order and the actual order, i.e., we $\textbf{minimize} ~ \varepsilon$
\begin{align}
&\textbf{subject to} \quad \left| N_{est}(i,j,k,l) - N(i,j,k,l) \right| < \varepsilon \notag \\
& \forall ~ i \in [1,2,...,I],~j \in [1,2,...,J],\notag \\
 &~~k \in [1,2,...,K],~ l \in [1,2,...,L].
\label{minimize}
\end{align}  
where $N_{est}(i,j,k,l)$ is given by
\begin{equation}
\label{nest_ijk}
N_{est}(i,j,k,l) = -\frac{\log_{10}(\bar{P}_\delta^{(l)})}{\Upsilon (\Delta_\omega^{(i)},W_r^{(j)})} + \Gamma (\Delta_\omega^{(i)},W_r^{(j)},\alpha^{(k)}),
\end{equation}
where $\Delta_\omega^{(i)}$, $W_r^{(j)}$, $\alpha^{(k)}$, and $P_\delta^{(l)}$ are discretized into $I$, $J$, $K$, and $L$ grid points within its given range, respectively. Here, $N(i,j,k,l)$ represents the actual filter order under the specific design parameters of $\Delta_\omega^{(i)}$, $W_r^{(j)}$, $\alpha^{(k)}$, and $P_\delta^{(l)}$, whereas $N_{est}(i,j,k,l)$ is calculated with the obtained ripples $\bar{\delta_p}$ and $\bar{\delta_s}$ as $\bar{P}_\delta^{(l)} = (\bar{\delta_p} \bar{\delta_s})^{(l)}$ and the design parameters of $\Delta_\omega^{(i)}$, $W_r^{(j)}$, and $\alpha^{(k)}$.

According to our observation based on numerous designs, the estimates for the $\Upsilon$ and $\Gamma$ functions are divided into two regions that are distinguished by the value of $W_r$, namely
\begin{equation}
\label{seg}
\begin{cases}
W_r \geq 1 & \text{Region}~1\\
W_r    < 1 & \text{Region}~2.
\end{cases}
\end{equation}
For Region 1, the $\Upsilon$ and $\Gamma$ are given, respectively, as
\begin{equation}
\label{beta}
\Upsilon = P_1 \times \Delta_\omega^{P_2} + P_3 \times \log_{10}(W_r) + P_4
\end{equation}
and
\begin{equation}
\label{gamma}
\Gamma = \left( \frac{Q_1}{\Delta_\omega} + Q_2 \right) \times \left( 1+ \log_{10}(W_r) \right)^{Q_3} + Q_4(\alpha-1) + Q_5. 
\end{equation}
For Region 2, the $W_r$ in the above functions of $\Upsilon$ and $\Gamma$ should be replaced by ${1}/{W_r}$. 
Further the best values of $P$ and $Q$ obtained in the minimax sense are given in Table \ref{tab:estpara}, where $\Delta_\omega^{(i)}$, $W_r^{(j)}$, $\alpha^{(k)}$, and $P_\delta^{(l)}$ are discretized into 10 grid points within the given range for the optimization. From the results, we can note that the order is largely dependent on $\Delta_\omega^{P_2}$ with an inversely proportional relation. This is similar to the order estimate of a regular FIR filter \cite{Rabiner_1975}, which however is independent of $\alpha$ and inversely proportional to $\Delta_\omega$. Hence, the derived order estimation formulas for the bandwidth extension filter are much more accurate than \cite{Rabiner_1975,Ichige_00}. This will be further illustrated in Section \ref{sec:simexa}.

\section{Design Examples}
\label{sec:simexa}
We first briefly discuss the design procedure for the examples.
1) With the given constraints of $\delta_p$ and $\delta_s$ as well as $\Delta_\omega$ and $\alpha$, calculate $W_r$ and then $N_{est}$ according to the derived estimates. 2) Design the bandwidth extension filter $h_r[n]$ of order $N_{est}$, and then search around $N_{est}$ to find the minimal order, say$N_{min}$, for which the given constraints are met.

\begin{figure}[t]
  \centerline{\includegraphics[width=7.7cm]{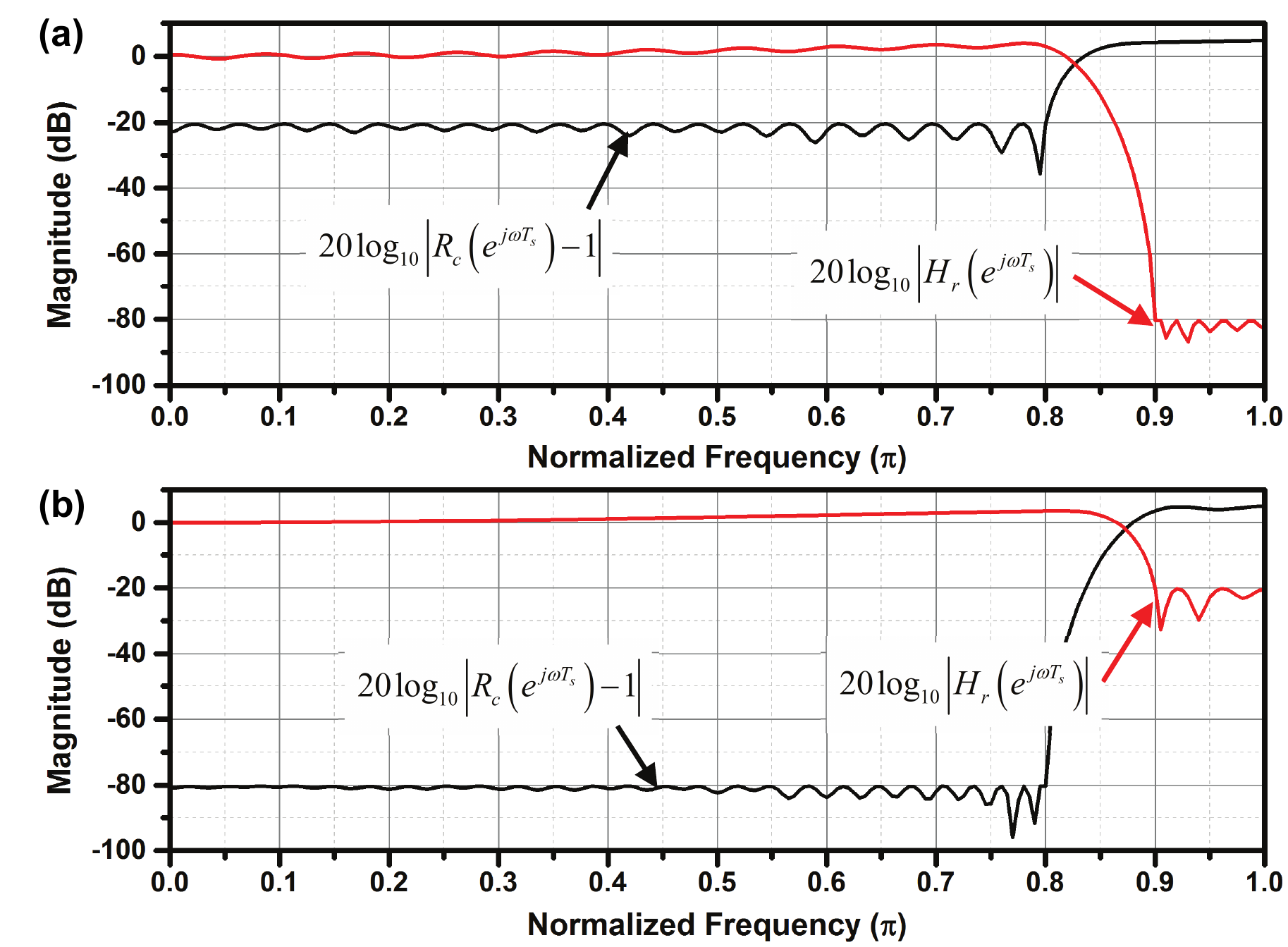}}  
  \caption{\footnotesize{Approximation-error modulus for the minimal-order filters. The black curves represent $20\log_{10} \vert R_c(e^{j\omega T_s}) -1\vert$, whereas the red curves show $20\log_{10} \vert H_r(e^{j\omega T_s})\vert$. (a) $N_{min}=48$ and $W_r=10^3$. (b) $N_{min}=57$ and $W_r=10^{-3}$.}}
  \label{fig:exam1}
\end{figure}

\textit{Example 1:} We configure $\omega_c T_s = 0.7 \pi$, $\omega_e T_s=0.8 \pi$, and $\Delta_\omega T_s = 0.1\pi$. First, the ripples constraints are set as $\delta_p = 0.1$ and $\delta_s = 0.0001$, thus $W_r=10^3$ that belongs to Region 1. Therefore, $N_{est}$ is calculated as $46.75$ and rounded to the nearest integer $47$. The bandwidth extension filter is designed in the minimax sense as presented in Section \ref{subsec:mini}, and we obtain the ripples as $\bar{\delta_p} = -19.16$~dB and $\bar{\delta_s} = -79.16$~dB. Since the given constraints are not satisfied, we increase the filter order to find $N_{min}$. One obtains $\bar{\delta_p} = -20.33$ dB and $\bar{\delta_s} = -80.33$ dB with the minimal order of 48, where the approximation errors are illustrated in Fig. \ref{fig:exam1}(a). Further we switch the ripple constraints as $\delta_p = 0.0001$ and $\delta_s = 0.1$. Then we have $N_{est} = 57.49$ by using the $P$ and $Q$ values for Region 2. The $h_r[n]$ is designed with $57$, and the performance achieves $\bar{\delta_p} = -80.23$ dB and $\bar{\delta_s} = -20.23$ dB. Here, the minimal order is $N_{min} = 57$, and the approximation errors are depicted in Fig. \ref{fig:exam1}(b). 

\begin{figure}[t]
  \centerline{\includegraphics[width=7.7cm]{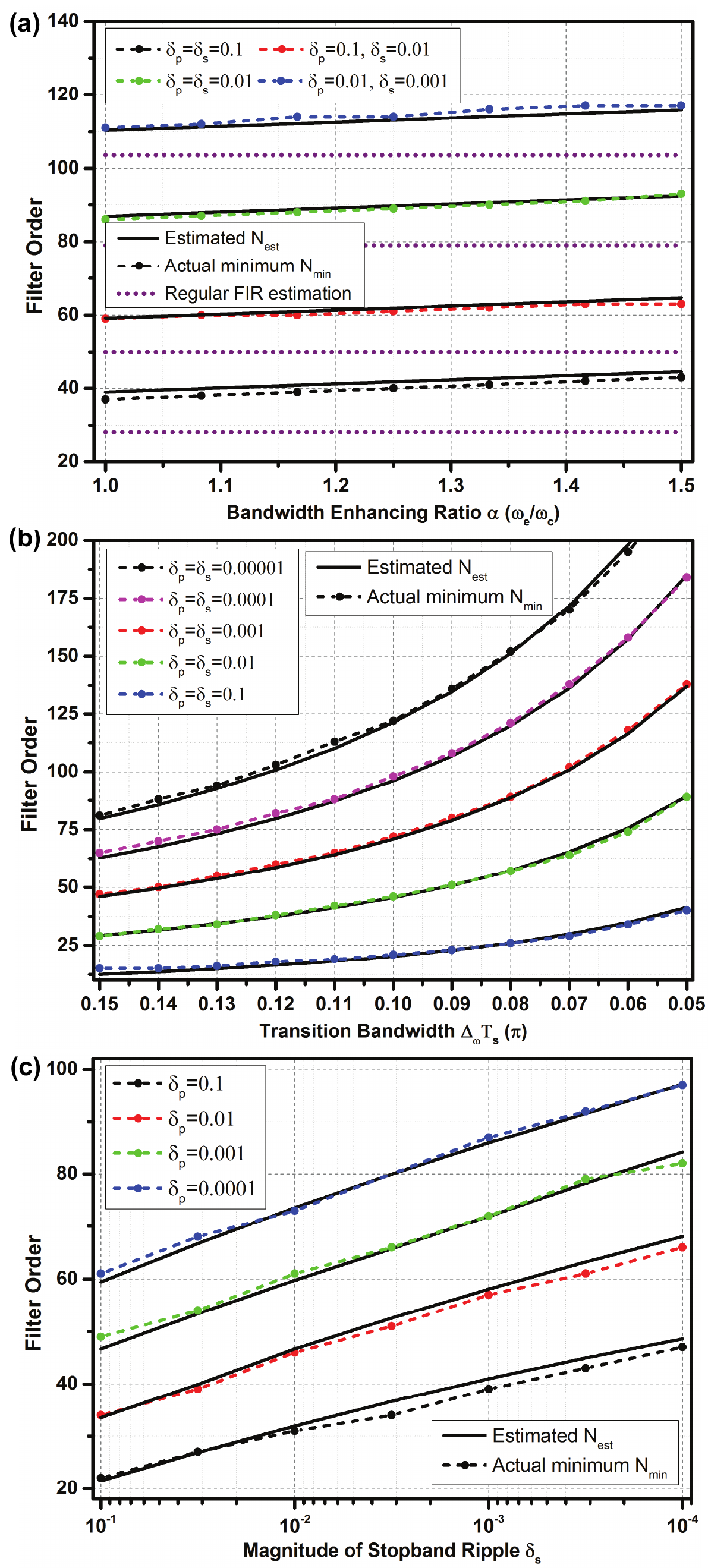}}  
  \caption{Estimation performance with various design parameters. The black solid and color dashed curves represent $N_{est}$ and $N_{min}$, respectively.}
  \label{fig:orderall}
\end{figure}

We can observe the necessity for dividing the $\Upsilon$ and $\Gamma$ functions into two regions, which is due to the unbalanced efforts paid on the passband and the stopband. Since the extended passband is generally far wider than the stopband, one needs to pay more complexity to reduce the passband ripple than the stopband ripple. Furthermore, if we design a regular FIR low-pass filter with a passband edge at $0.8\pi$ and transition band of $0.1\pi$, to achieve the same ripple performance as in Fig. \ref{fig:exam1}(a) and (b), the orders of $42$ and $53$ are required, respectively. Hence, one has to pay more complexity to extend the bandwidth than to design a regular FIR filter, which means that the order estimations available for the regular FIR filters, \cite{Rabiner_1975,Ichige_00}, are not accurate for the bandwidth extension filter. In Fig. \ref{fig:orderall}(a), we also indicate the necessity of the derived order estimates for the bandwidth extension filter, where the purple dotted lines indicate the estimations for regular FIR filters as in \cite{Rabiner_1975} with the same ripple requirements and transition bandwidth.

\textit{Example 2:} To comprehensively validate the proposed filter order estimation, we verify its performance with various $\delta_p$, $\delta_s$, $\Delta_\omega$, and $\alpha$, which is graphically exemplified in Fig. \ref{fig:orderall}. 
First we verify the estimation accuracy for different expansion ratios as depicted in Fig. \ref{fig:orderall}(a), where $\Delta_\omega T_s = 0.05\pi$. In Fig. \ref{fig:orderall}(b), we show the performance with different transition bandwidths, whereas other parameters are set as $\omega_c T_s = 0.66 \pi$ and $\omega_e T_s=0.8 \pi$. Furthermore, we simulate with various $\delta_p$ and $\delta_s$ as shown in Fig. \ref{fig:orderall}(c), where we configure $\omega_c T_s= 0.65 \pi$, $\omega_e T_s=0.85 \pi$, and $\Delta_\omega T_s = 0.1\pi$. From Fig. \ref{fig:orderall}, it is observed that the estimated order $N_{est}$ matches well with the actual minimal order $N_{min}$ under various conditions.

\section{Conclusions}
\label{sec:conclusions}
A bandwidth extension method for ADCs, utilizing FIR filters designed in the minimax sense, was proposed. We derived the estimation of the order requirement to meet a given specification in terms of $P_\delta$, $\Delta_\omega$, $W_r$, and $\alpha$. Simulation results indicate the estimation accuracy. With the derived formulas, one can achieve an accurate complexity assessment at the top-level design of the overall systems and conveniently find the minimal order that helps in reducing the design time.




\section{Acknowledgment}
This work was supported in part by a project within the Swedish strategic research center Security Link.


\end{document}